\documentclass[aps,prb,epsfig,preprint,superscriptaddress,showpacs]{revtex4-1}
\usepackage{epstopdf}
\usepackage{graphicx}  
\usepackage{dcolumn}   
\usepackage{bm}        
\usepackage{amssymb}
\usepackage[english]{babel}

\begin{document}

\title{Anomalous structural behavior and antiferroelectricity in BiGdO$_{3}$: Detailed temperature and high pressure study}

\author{Rajesh Jana}
\email [Corresponding Author: ]{rajeshjana8@gmail.com}
\affiliation{Department of Physical Sciences, Indian Institute of Science Education and Research Kolkata, Mohanpur Campus, Mohanpur 741246, Nadia, West Bengal, India.}
\affiliation{
 Beijing National Laboratory for Condensed Matter Physics and Institute of Physics, Chinese Academy of Sciences, Beijing 100190, China
}%
\author{Apurba Dutta}
\affiliation{CMP Division, Saha Institute of Nuclear Physics, HBNI, 1/AF, Bidhannagar, Kolkata-700064, India}
\author{Pinku Saha}
\affiliation{Department of Physical Sciences, Indian Institute of Science Education and Research Kolkata, Mohanpur Campus, Mohanpur 741246, Nadia, West Bengal, India.}
\author{Kapil Mandal}
\affiliation{Department of Physical Sciences, Indian Institute of Science Education and Research Kolkata, Mohanpur Campus, Mohanpur 741246, Nadia, West Bengal, India.}
\author{Bishnupada Ghosh}
\affiliation{Department of Physical Sciences, Indian Institute of Science Education and Research Kolkata, Mohanpur Campus, Mohanpur 741246, Nadia, West Bengal, India.}
\author{Amreesh Chandra}
\affiliation{Department of Physics, Indian Institute of Technology Kharagpur, Kharagpur-721302, India}
\author{I. Das}
\affiliation{CMP Division, Saha Institute of Nuclear Physics, HBNI, 1/AF, Bidhannagar, Kolkata-700064, India}
\author{Goutam Dev Mukherjee}
\email [Corresponding Author: ]{goutamdev@iiserkol.ac.in}
\affiliation{Department of Physical Sciences, Indian Institute of Science Education and Research Kolkata, Mohanpur Campus, Mohanpur 741246, Nadia, West Bengal, India.}

\begin{abstract}
A detailed temperature and pressure investigation on BiGdO$_{3}$ is carried out by means of dielectric constant, piezoelectric current, polarization-electric field loop, Raman scattering and x-ray diffraction measurements. Temperature dependent dielectric constant and dielectric loss show two anomalies at about 290 K (T$_r$) and 720 K (T$_C$).  The later anomaly is most likely due to antiferroelectric to paraelectric transition as hinted by  piezoelectric current and polarization-electric field loop measurements at room temperature, while the former anomaly suggests reorientation of polarization. Cubic to orthorhombic structural transition is observed at about 10 GPa in high pressure x-ray diffraction studies accompanied by anisotropic lattice parameter changes. An expansion about 30 \% along $a$-axis and 15 \% contraction along $b$-axis during the structural transition result in 9.5 \% expansion in unit cell volume. This structural transition is corroborated by anomalous softening and large increase in full width half maximum (FWHM) of 640 cm$^{-1}$ Raman mode above 10 GPa. Enhancement of large structural distortion and significant volume expansion during the structural transition indicate towards an antiferroelectric to ferroelectric transition in the system. 
 
\end{abstract}

\pacs{77.80.-e,  62.50.-p,  64.70.Nd }
\maketitle

\section{Introduction}

	Ferroelectricity is the property of a certain class of materials which exhibit reversible spontaneous polarization in the presence of an electric field.  Since its discovery ferroelectric (FE) materials have been extensively studied for their novel properties viz. spontaneous polarization, piezoelectricity, pyroelectricity, coupling with magnetic order (multiferroic), etc. On the other hand, antiferroelectric materials are less studied compared to FE materials. Using a phenomenological approach, C. Kittel proposed antiferroelectric (AFE) behaviour in 1951 \cite{Kittel}. An AFE state can be defined as the lines of ions in which two neighbouring lines are oppositely polarized. Therefore, there will be no net polarization even with the  application of the electric field and they will not be piezoelectric in contrast to the ferroelectric materials. However, a certain high electric field (E$_F$) will force the antiparallel dipoles to reorient in parallel fashion  and an electric  field induced AFE to FE transition can be achieved since both the FE and AFE phases have similar free energy. Shirane \textit{et al.} for the first time reported the AFE behavior in PbZrO$_{3}$ below 230 K \cite{Shirane}. Thereafter, antiferroelectricity was observed in PbHfO$_{3}$ and in other materials - NaNbO$_{3}$ , AgNbO$_{3}$, Pb(Mg$_{0.5}$W$_{0.5}$)O$_{3}$, Pb(Yb$_{0.5}$Nb$_{0.5}$)O$_{3}$ \cite{Shirane2, Shirane3, Francobe, Smolenskii, Yamamoto}.   In recent times antiferroelectric materials have generated considerable renewed research interest owing to their promising ultrahigh energy density, large strain, giant electrocaloric effect \cite{Zhang, Chauhan, Bao, Zhang2, Zhao, Tian, Liu, Pirc} etc. AFE materials also possess some indispensable properties for industrial applications such as - low dielectric loss, low coercive field, low remanant polarization, and fast discharge rates. 

	Bi-based perovskite oxides BiMO$_{3}$ where M is a transition metal ion have drawn special research interest in designing new multiferroics. In these materials ferroelectricity arises due to  6s lone pair electrons of  Bi$^{3+}$ ion which shift towards oxygen octahedra to form local dipole structure. On the other hand, presence of transition metal ion with partially filled d shell at the B site makes long range magnetic ordering. Multiferroic properties have been observed several FE materials such as - BiFeO$_{3}$, BiCrO$_{3}$, BiNiO$_{3}$, BiCoO$_{3}$ and BiMnO$_{3}$ \cite{Kiselev, Niitaka, Cai, Belik, Kimura} etc. Besides ferroelectricity, AFE behavior has also been evidenced in some Bi based oxides. For example, Kim \textit{et al.}\cite{Kim} and Kan {et al.}\cite{Kan} reported AFE signature in BiCrO$_{3}$ thin films  and rare‐earth doped BiFeO$_{3}$ respectively. Baettig \textit{et al.} theoretically predicted antiferroelectric arrangement in centrosymmetric BiMnO$_{3}$ originated from stereo-chemical activity of Bi$^{3+}$ 6s$^{2}$ lone pair electrons \cite{Baettig}.  In BiFeO$_{3}$ and BiMnO$_{3}$ two outer 6s electrons of Bi ion, instead of making chemical bond, shift away from Bi ion towards oxygen octahedra and produce local dipoles which may align in ferroelectric or antiferroelectric arrangement \cite{Khomskii, Seshadri, Ravindran}. In a recent study, Saha \textit{et al.} suggested dielectric relaxation and antiferromagnetic interaction in BiGdO$_{3}$ (BG) \cite{Saha}. However, as per authors knowledge no detailed study on structural and electronic properties of this material is available in literature. In this context we have synthesized and carried out a detailed investigation on BiGdO$_{3}$ by means of dielectric constant, piezoelectric current, polarization-electric field loop, Raman scattering and x-ray diffraction measurements at ambient as well as high pressures. Two anomalies have been observed in the temperature dependent dielectric constant profile of BG at about 290 K and 720 K. Piezoelectric current and polarization-electric field loop measurements at room temperature indicate antiferroelectric behaviour in BG below 720 K. High pressure x-ray diffraction (XRD) and Raman studies show cubic to orthorhombic structural transition at 10 GPa accompanied by a large  volume expansion. We speculate that pressure induced large increase in structural distortion and significant volume expansion may lead to AFE to FE transition.

\section{experiment }
	Polycrystalline BG sample is synthesized using conventional solid-state reaction method. Stoichiometric amount of 99.99 \% pure (Sigma Aldrich) Bi$_{2}$O$_{3}$ and Gd$_{2}$O$_{3}$ are thoroughly mixed using agate mortar and pestle for several hours in 2-propanol medium. Pellets of well mixed powder are then  calcined at 930$^{0}$C for about 9h. This calcination process is repeated  to improve crystalline quality of the sample. Formation of pure single phase of the sample is confirmed by the collecting x-ray diffraction pattern using Rigaku SmartLab x-ray diffractometer with 9 KW rotating anode and Cu K$_{\alpha}$ x-ray source. Temperature dependent dielectric constant measurements are carried out on sintered BG pellets of 8 mm \textit{dia} and 2 mm thickness after polishing both the faces of the pellets with 0.25 $\mu$m  silver paste.  It is first washed with acetone to clean its surfaces and then dried at 120$^{0}$ for 20 min to remove moisture. A N4L PSM 1735 impedance analyzer is used to explore temperature dependence dielectric constant in the frequency range 100 Hz to 1 MHz. Temperature is calibrated using a K type thermocouple and a digital voltmeter.  Piezoelectric current measurements are carried out using toroid anvil cell and 300 ton hydraulic press. Detailed experimental procedure for this measurement can be found elsewhere \cite{Jana, Jana2}.  Room temperature polarization-electric field hysteresis loop measurements are carried out using Radiant Technology’s Precision 10 KV HVI-SC loop measurement system. High pressure x-ray diffraction and Raman spectroscopic measurements are carried out using a piston type diamond anvil cell (easyLab Co., UK) with a culet size of 300 $\mu$m . The sample along with pressure marker are loaded in the 100 $\mu$m  central hole  of a pre-indented steel gasket. A mixture of methanol and ethanol is also added in the 4:1 ratio in order to maintain hydrostatic condition inside the sample chamber. High pressure Raman spectra are taken using Horiba Jobin-Yvon Lab RAM HR-800 spectrometer with 1800 g/mm grating. The samples are excited with 488 nm line of Ar-ion laser and pressure inside the sample chamber are determined  using Ruby fluorescence technique \cite{Mao}. XRD measurements at high pressures are carried out in EXPRESS beamline in the Elettra synchrotron light source, Italy, using a monochromatic wavelength 0.5 \AA. In XRD studies, pressure is determined from isothermal P-V equation of sate of silver \cite{Dewaele}, small quantities of which was mixed with sample before loading in diamond anvil cell. X-ray beam is collimated to about 30 $\mu$m and diffracted x-rays are collected using a MAR 3450 image plate detector aligned normal to the incident beam. Collected two dimensional patterns are converted to intensity 
versus 2$\theta$ profile using FIT2D software \cite{Hammersley}. Full profile fitting of the XRD patterns are carried out using GSAS \cite{Toby} after indexing using the software CRYSFIRE \cite{Shirley}.

\section{results and discussions}


	Temperature dependence of dielectric constant and dielectric loss of BG in the range 77 K to 755 K at a few selected frequencies are shown in Fig. 1. Dielectric constant and dielectric loss show  two anomalies in their temperature profiles. On increasing temperature from 77 K, dielectric constant slowly increases and exhibit a broad peak around 290 K (T$_r$) for all  frequencies in the range 10 KHz to 1 MHz.  At this temperature a peak in the dielectric loss has also been observed at all frequencies except for 1 MHZ frequency. Upon further increasing temperature, dielectric constant start to increase rapidly above 550 K with a clear peak at about 720 K (T$_C$). A strong frequency dispersion in the dielectric constant is observed. The absence of peak position dependence on frequency indicates non-relaxor type behaviour which is in contrast to the previous literature report describing BG as a relaxor type ferroelectric \cite{Saha}. Temperature dependent dielectric loss also increases sharply around T$_C$. Above anomalous behavior in dielectric constant and dielectric loss around T$_C$ are the signatures of ferro/antiferroelectric to paraelectric  phase transition while the anomaly around T$_r$ could be  due to reorientation of polarization. However, x-ray diffraction analysis establish that BG crystallizes in Fm-3m centrosymmetric cubic structure. The structural symmetry does not allow it to be a conventional ferroelectric material. In order to check further about the presence of ferroelectric order we have carried out piezoelectric current measurements at room temperature (298 K)  using TA apparatus \cite{Jana, Jana2}. The obtained data are plotted in Fig. 2(a), which shows a small fluctuating current even with reversing the direction of electric field. A ferroelectric material should show piezoelectric effect and the polarity of piezoelectrically induced voltage must be reversed with reversal of applied electric field. In this case we do not observe any polarity change with change in applied poling field direction indicating the absence of ferroelectric order in the sample. Similar sudden increase in dielectric constant near the transition temperature and non-relaxor type of  behavior have been observed in several antiferroelectric materials like – PbZrO$_{3}$, NaNbO$_{3}$, PbHfO$_{3}$,  Cs$_{2}$Nb$_{4}$O$_{11}$ \cite{Shirane, Shirane3, Samara, Smith, Liu2}. Therefore , one can speculate that BG may be antiferroelectric below 720 K. Electric field (E) dependent polarization (P) measurement is the best experimental way to confirm ferro/antiferroelectric behaviour in a material. We have carried out room temperature polarization-electric field loop measurements and the corresponding data are shown in Fig. 2(b). A ferroelectric material should exhibit saturation in polarization at high electric field. However, in our study we have observed unsaturated elliptical loop with very small area up to the highest applied field of 50 KV/cm which exhibits lossy capacitor behaviour \cite{Evans}. We could not apply higher field above 50 KV/cm due to dielectric breakdown.  The lossy dielectric response can be explained by considering the phase difference between polarization and applied electric field. For an ideal linear capacitor, polarization and electric field are in phase and one would get a linear behaviour in polarization-electric field loop. On the other hand an ideal resistor produces circular polarization-electric field loop as current and voltage are in phase. Therefore, when linear capacitor and ideal resistor are connected in parallel combination, it gives lossy dielectric behaviour with elliptical loop.  In several materials antiferroelectric behaviour is confirmed by the field induced double hysteresis loop under high electric field. However, in our study, we do not observe this type of behaviour. Probably the critical value for electric field (E$_C$) to induce AFE to FE transition is greater than the breakdown fields at room temperature. Therefore, our experimental observations, such as, peak in the temperature dependent dielectric constant, non-appearance of piezoelectricity and absence of hysteresis nature in polarization-electric field loop suggest that the sample is most likely antiferroelectric below 720 K.


	In order to investigate structural behavior of this sample at high pressures, we have conducted anle dispersive x-ray diffraction measurements under pressure. Ambient and a few selected XRD patterns with increasing pressure up to 22.5 GPa are shown in Fig. 3. Ambient XRD analysis of BiGdO$_3$ shows a cubic Fm-3m phase having lattice parameters of 5.47551(4) \AA. We have not observed any remarkable changes in the diffraction patterns except peak broadening up to 6.4 GPa. As the pressure increases Bragg peaks shift to higher 2$\theta$ values due to lattice contraction. At about 7.2 GPa onset of a new Bragg peak around 2$\theta$ = 10º in between first two reflection lines is observed and it  becomes  prominent at about 10 GPa as shown by the vertical arrows in Fig. 3. On further increase of  pressure the intensity of the new peak gradually increases while the intensity of its two neighbouring lines decreases rapidly and these three peaks finally merge at about 18 GPa. We do not observe any further change other than broadening of all the peaks up to 22.5 GPa. The XRD pattern above 9 GPa could not be fitted with ambient Fm-3m cubic structure. Therefore, the appearance of the new reflection line at 10 GPa definitely indicate towards a structural transition in the sample. On indexing, the high pressure phase is found to be  orthorhombic  with Pbca symmetry having lattice parameters a = 6.87935(2) \AA, b = 5.30753(4) \AA and c= 4.48219(3) \AA.  We have carried out full structural refinement using Rietveld refinement process taking starting atomic coordinates from similar structure material \cite{Redhammer}.  Rietveld refined XRD pattern of BG with Pbca structure at 12.1 GPa and 16.5 GPa are shown in Fig. 4 (b) and (c). The obtained structural parameters for 12.1 GPa XRD data are presented in Table I. Fig.  5 shows the pressure evolution of unit cell lattice parameters  for both  cubic and orthorhombic structures. In the Fig. 5(a) one can see that structural transition is accompanied by large expansion (about 30 \%) along $a$-axis, 15 \% contraction along $c$-axis, while no significant change is observed along $b$-axis. As a result of this, unit cell volume is expanded by about 9.5 \% (Fig. 5(b)). In the inset of Fig. 5(b) a gradual decrease in the c/a ratio in orthorhombic phase is exhibited. P-V data in both the phases could be fitted with 3rd order Birch-Murnaghan equation of state \cite{Birch, Murnaghan} 

\begin{equation}
P = {3/2}B_{0}[(V_{0}/V)^{7/3} - (V_{0}/V)^{5/3}]\times[1-3/4(4-B')\{(V_0/V)^{2/3}-1\}]
\end{equation}

Where, B$_{0}$ and V$_{0}$ are the bulk modulus and volume at ambient pressure respectively. B$^\prime$ is the first order pressure derivative of bulk modulus. For the first region up-to 9 GPa our best fit gives V$_{0}$ = 164.29(2) \AA$^{3}$, B$_{0}$ = 67.42(3) GPa, and B$^\prime$ = 4.99(6). In high pressure region these values become V$_{0}$ = 184.61(7) \AA$^{3}$, B$_{0}$ = 66.14(5) GPa, and B$^\prime$ = 4.69(6). From the calculated bulk modulus it is established that despite large volume expansion the compressibility of the sample remains almost same after cubic to orthorhombic transition.


	As a complementary study, we have carried out high pressure Raman studies on BG to  probe structural behavior more deeply. Factor group analysis for BiGdO$_{3}$ with Fm-3m structure predicts five (A$_{1g}$ + E$_{g}$  + 3T$_{2g}$) Raman active modes associated with oxygen vibration  \cite{Rousseau, Saha} .  High pressure Raman spectra at selective pressure values are shown in Fig. 6. At ambient pressure we have observed six Raman modes at 104, 140, 241, 368, 586 and 640 cm$^{-1}$. The observed Raman modes are found to be at higher values than the previously reported values \cite{Saha}. One reason could be that high quality sample is used in the present study, whereas a trace of secondary phase was present in the earlier study \cite{Saha}.  On increasing pressure two prominent Raman modes 104 cm$^{-1}$ (M$_1$) and 640 cm$^{-1}$ (M$_2$)  harden up to 10 GPa. Then the high frequency M$_2$ mode starts softening  above 10 GPa where the systems transform to orthorhombic structure as observed in high pressure x-ray diffraction studies. Upon further increase in pressure the intensity of M$_1$ mode rapidly decreases and broadening in the M$_2$ mode significantly increases. In Fig. 7(a), we have shown high pressure behavior of M$_2$ Raman mode. Upon increasing pressure it linearly increases with a slope of 3.58(0.06) cm$^{-1}$/GPa up to 10 GPa followed by a drastic fall which is corroborated by the sudden volume expansion while system transform to orthorhombic phase around this pressure. Above 15 GPa this mode frequency again increases linearly up to 18.3 GPa with a slope 2.94(0.18) cm$^{-1}$/GPa. At higher pressures error in mode fitting parameters increases drastically due to peak broadening. Emergence of large octahedral distortion is also reflected with rapid increase in the full width at half-maximum (FWHM) of the M$_2$ mode above 10 GPa (Fig. 7(b)). At about 25 GPa M$_1$ mode is found to have disappeared  completely.


	Shirane {et al.} and Swaguchi \textit{et al.} did not observe any electric field induced  AFE to FE transition in the first reported antiferroelectric  PbZrO$_{3}$ in its polycrystalline form at room temperature \cite{Shirane, Sawaguchi}. The double hysteresis behavior was only seen near the Curie temperature (533 K) and using a linear extrapolation of  threshold electric field (E$_C$), it was calculated that 360 KV/cm field is needed to trigger the force field transition at 298 K in PbZrO$_{3}$ polycrystal \cite{Shirane, Sawaguchi, Tan}. Since the value of E$_C$ is larger than the dielectric breakdown strength, field induced AFE to FE transition can not be observed at room temperature in ceramic PbZrO$_{3}$ sample. Later, Fesenko \textit{et al.} was able to detect the AFE to FE transition in the broad range of temperature (103 K to 433 K) in PbZrO$_{3}$ single crystal \cite{Fesenko}. In NaNBO$_{3}$, double hysteresis feature was also observed in only single crystal sample and E$_C$ is found to increase with decreasing temperature \cite{Tan, Cross}. Thus, polycrystalline antiferroelectric materials may or may not exhibit field induced transition at room temperature. Most probably this is the reason behind non observation of double hysteresis loop in our polycrystalline sample. With appropriate doping or making a thin film  may reduce the E$_C$ \cite{Cross2}.  To explain the presence of antiferroelectricity and transition to paraelectric state in cubic BiGdO$_{3}$, we need to discuss about the lone pair mechanism.  Analyzing crystal chemistry, Volkova \textit{et al.} showed that presence of lone pair in Bi$^{3+}$ compound only confirm the local distortion in cation polyhedra and emergence ferroelectricity requires high degree of stereo-chemical activity of the one pair which increases with decreasing temperatures \cite{Volkova}. They have also suggested that external perturbation of temperature and pressure will strongly affect the lone pair activity by changing its shape and position. Hence, one can anticipate that an antiferroelectric/ferroelectric to paraelectric transition in Bi$^{3+}$ compound at high temperatures. This same mechanism is applicable for reorientation of polarization at T$_r$.  Ram Seshadri and Nicola A. Hill  reported that ferroelectricity and highly distorted monoclinic structure in BiMnO$_{3}$ arises due to lone pair activity \cite{Seshadri}. Lone pair driven ferroelectricity/antiferroelectricity were also observed in  other materials - AgSbSe$_{2}$, BiAlO$_{3}$, Pb$_{2}$MgWO$_{6}$ \cite{Aggarwal, Kaczkowski, Seshadri2}. In a recent study, pair distribution function (PDF) analysis revealed that stereo-chemical activity of the 6s$^{2}$ lone pair  produces signiﬁcant lattice anharmonicity and local distortion in AgBiS$_{2}$ while on average the system showed cubic Fm-3m structure \cite{Rathore}.  This local distortion can be detected by local probe like Raman spectroscopy. Appearance of significant asymmetric Raman modes suggest that the local distortions are indeed present in cubic BiGdO$_{3}$. Bao et. al. have shown that La doping in Pb(Yb$_{0.5}$Nb$_{0.5}$)$_{0.92}$Ti$_{0.8}$O$_{3}$ compound increased c/a ratio and reduced [BO$_{6}$] octahedral distortion which in turn stabilize the AFE phase \cite{Bao}. Conversely it might  happen that decrease in $c/a$ ratio will stabilize FE phase increasing [BO$_{6}$] octahedral distortion. In our HP XRD study c/a ratio has been observed much less than 1 in the orthorhombic phase, which further decreases with pressure. Raman measurements also revealed large enhancement of structural distortion as evidenced from large increase in FWHM of M$_2$ Raman mode above 10 GPa. Field induced AFE-FE transition generally accompanied  by large increase in polarization and unit cell volume \cite{Chauhan, Liu, Sawaguchi, Pan}. In our study 9.5 \% volume expansion might be related to antiferroelectric to ferroelectric transition due to pressure induced large structural distortion.

\section{Conclusion}

We have synthesized and studied detailed structural and electronic behaviour of BiGdO$_{3}$. Temperature dependent dielectric constant and dielectric loss show two anomalies at about 290 K (T$_{r}$) and 720 K (T$_{m}$). The later anomaly most likely due to antiferroelectric to paraelectric transition as hinted by  piezoelectric current and room temperature polarization-electric field loop measurements, while the former anomaly suggest reorientation of polarization. We attributed both anomalies to the change in degree of  stereochemical activity of the Bi$^{3+}$ lone pair with increasing temperatures. Cubic to orthorhombic structural transition has been observed at about 10 GPa in high pressure x-ray diffraction studies accompanied by anisotropic lattice parameters changes. An expansion about 30 \% along a axis and 15 \% contraction along b axis during the structural transition result  9.5 \% expansion in the unit cell volume. This structural transition is corroborated by anomalous softening of the high wave number Raman mode frequency. Enhancement of structural distortion and large volume expansion indicate towards an antiferroelectric to ferroelectric transition above 10 GPa. Further studies are needed to understand ferroelectric behaviour if any, at high pressures.

\section{Acknowledgments}
The authors gratefully acknowledge the financial support through project grant MoES/16/25/10-RDEAS
from Ministry of Earth Sciences, Government of India. The authors also acknowledge Department of Science and Technology (DST), Indo-Italian Program of Cooperation and Elettra Synchrotron Light Source for financial and laboratory support for synchrotron based x–ray diffraction measurements. 

\newpage  
\begin{table}[b]
\caption{\label{tab:table2}
Rietveld refined crystal structural parameters of BiGdO$_{3}$ obtained from x-ray diffraction data at 12.1 GPa 
in space group Pbca (No. 61) with lattice parameters a = 6.8642(7) \AA, b = 5.3113(6) \AA, c = 4.4578(6) \AA, and V = 162.522(34) \AA$^{3}$; Goodness of ﬁt: $\chi^2$ = 1.49, wR$_p$ = 3.01\%, R$_p$ 2.9\%}
\begin{ruledtabular}
\begin{tabular}{cccccc}
 atom & site& x & y & z & U$_{iso}$ \\
\hline
Bi1& 8c & 0.56991(2)  & 0.94666(2)  & 0.92461(4) & 0.04696(3)\\
Bi2& 8c & 0.37076(7)  & 0.41722(9)  & 0.38707(1) & 0.05351(4)\\
Gd1& 8c & 0.31035(7)  & 0.32019(8)  & 0.17058(3) & 0.05075(4)\\
Gd2& 8c & 0.32073(5)  & 0.45330(3)  & 0.76125(2) & 0.04806(1)\\
O1& 8c & 0.18480(12)  & 0.33960(13) & 0.03870(13) & 0.02543(2)\\
O2& 8c & 0.31180(17)  & 0.49640(18) & 0.05820(14) & 0.02440(4)\\
O3& 8c & 0.30250(14)  & 0.23630(16) & 0.81630(19) & 0.02648(6)\\
O4& 8c & 0.56110(21)  & 0.33650(24) & 0.78680(12) & 0.02552(6)\\
O5& 8c & 0.43320(16)  & 0.48060(17) & 0.69320(18) & 0.02495(4)\\
O6& 8c & 0.44760(26)  & 0.20280(29) & 0.58680(21) & 0.02582(2)\\
\end{tabular}
\end{ruledtabular}
\end{table}
 
\begin{figure}
\includegraphics[width = 8cm]{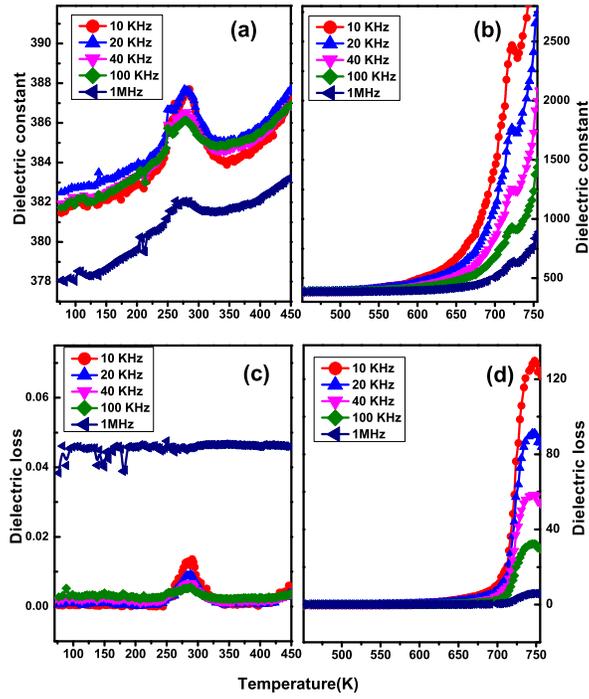}
\caption{\label{fig1}(Colour online) Temperature dependent dielectric constant and dielectric loss at a few selected frequencies in the temperature range 77K - 450K shown in (a) and (c); range 450K - 755K in (b) and (d)}
\end{figure}

\begin{figure}
\includegraphics[width=8cm]{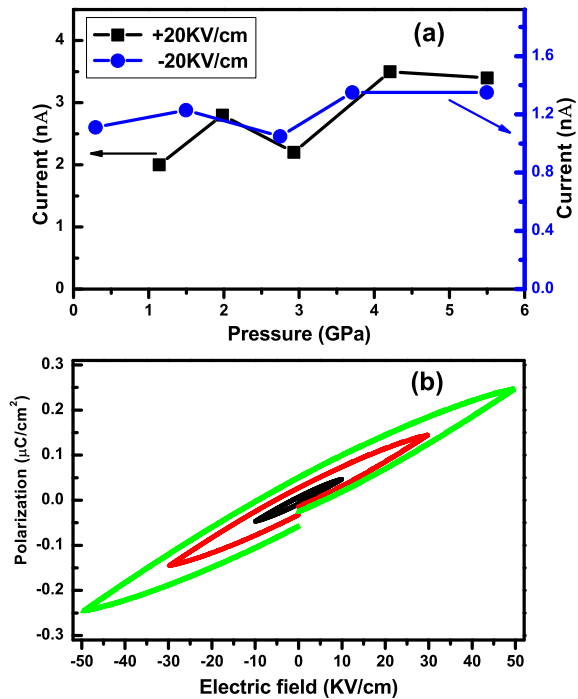}
\caption{\label{fig2}(Colour online) (a) Piezoelectric current measured with reducing pressure after poling the sample under positive and negative electric field of 20KV/cm at 5.5 GPa. (b) Polarization-electric field measurements up to 50KV/cm at room temperature and ambient pressure.}
\end{figure}  

\begin{figure}
\includegraphics[width=8cm]{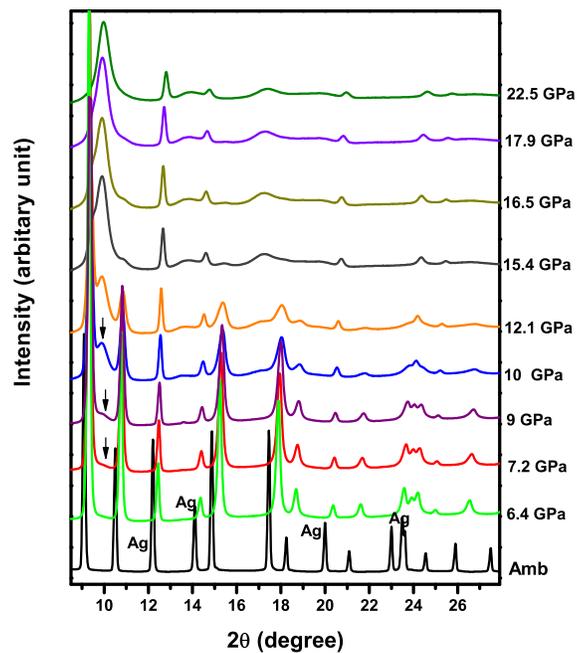}
\caption{\label{fig3}(Colour online) High pressure XRD spectra (x-ray wavelength = 0.5 Å) at selected pressures along with pressure calibrate silver peaks which are marked by Ag.  Appearance of a new Bragg reflection line is observed at 10 GPA (Shown by arrow).}
\end{figure}  
      
\begin{figure}
\includegraphics[width=8cm]{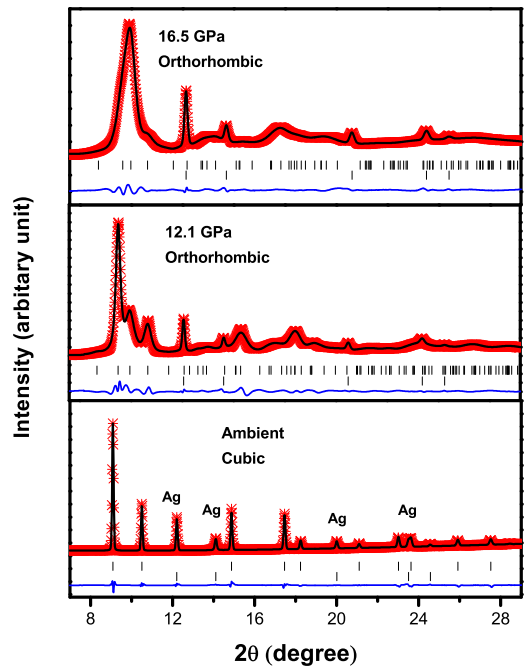}
\caption{\label{fig4}(Colour online) (a) Rietveld  refined XRD pattern of BiGdO$_3$ at ambient pressure, 12.1 GPa and 16.5 GPa. The red stars are the observed data points, the black line shows the fit to the data, Bragg positions of the sample are marked with vertical ticks and the Bragg peaks of Ag marked by the ticks at a slightly lower level, and the difference is marked by the blue line.} 
\end{figure}

\begin{figure} 
\includegraphics[width=8cm]{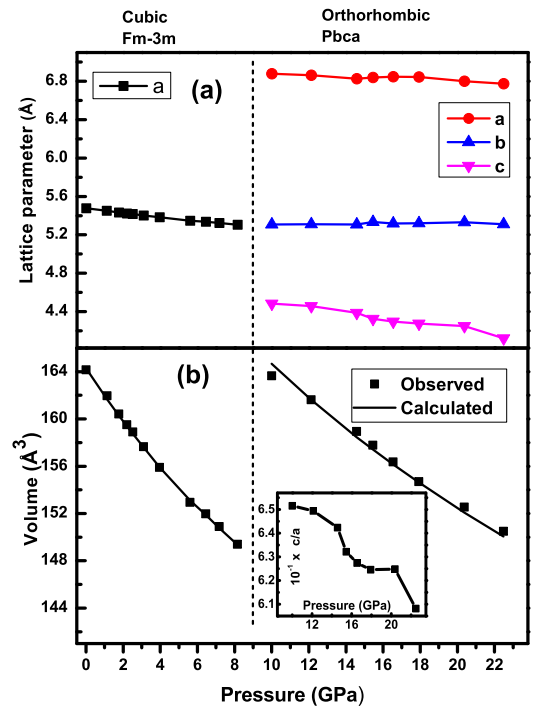}
\caption{\label{fig5}(Colour online) (a) Pressure evolution of unit cell lattice parameters before and after transition in  cubic and orthorhombic structure respectively. (b) Pressure-volume data fitted to two different 3rd order Birch-Murnaghan equation of state for two different regions of pressure. Inset figure shows c/a ratio with pressure in the orthorhombic phase.}
\end{figure}    
 \begin{figure}
\includegraphics[width=8cm]{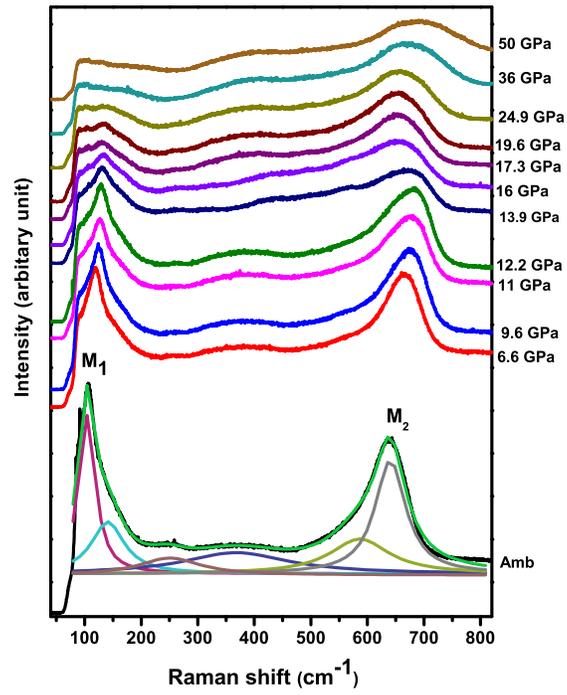}
\caption{\label{fig6}(Colour online)
High pressure Raman spectra at selected pressure values up to 50 GPa. Deconvolution of six Raman modes shown for ambient pressure data}
\end{figure}

  \begin{figure}
\includegraphics[width=8cm]{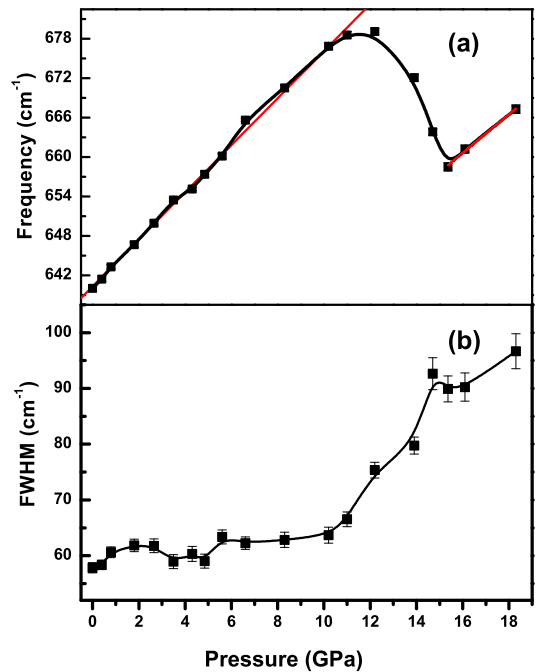}
\caption{\label{fig7}(Colour online)
 Pressure evolution of 640 cm$^{-1}$ (M$_2$) mode (a) frequency; and (b) FWHM with error bar up to 18.3 GPa. Error bar is less than data point symbol for the mode frequency. Red line in the figure (a) is the linear fitting with slope m = 3.58(0.06) cm$^{-1}$/GPa below 10 GPa and m = 2.94 (0.18) cm$^{-1}$/GPa above 15 GPa } 
 \end{figure}

\end{document}